\documentclass[10pt,letterpaper]{article}
\usepackage{opex3}

\usepackage{graphicx}

\begin{document}

\title{Field Test of Classical Symmetric Encryption with Continuous Variable Quantum Key Distribution}

\author{Paul~Jouguet$^{1,2}$, S{\'e}bastien~Kunz-Jacques$^1$, Thierry~Debuisschert$^3$, Simon~Fossier$^{3}$, Eleni~Diamanti$^2$, Romain~All{\'e}aume$^2$, Rosa~Tualle-Brouri$^4$, Philippe~Grangier$^4$, Anthony~Leverrier$^5$, Philippe~Pache$^6$, and Philippe~Painchault$^6$}
\address{$^1$SeQureNet, 23 avenue d'Italie, 75013 Paris, France}
\address{$^2$Institut Telecom / Telecom ParisTech, CNRS LTCI, 46 rue Barrault, 75634 Paris Cedex 13, France}
\address{$^3$Thales Research \& Technology France, 1 av. Augustin Fresnel, Campus Polytechnique, 91767 Palaiseau Cedex, France}
\address{$^4$Laboratoire Charles Fabry, Institut d'Optique - CNRS - Univ. Paris-Sud 11, 2 av. Augustin Fresnel, 91127 Palaiseau Cedex, France}
\address{$^5$ICFO - The Institute of Photonic Sciences Av. Carl Friedrich Gauss, num. 3 E-08860 Castelldefels (Barcelona), Spain}
\address{$^6$Thales Communications S.A., 160 boulevard de Valmy, BP 82, 92704 Colombes Cedex}
\email{paul.jouguet@telecom-paristech.fr}

\begin{abstract}
We report on the design and performance of a point-to-point classical symmetric encryption link with fast key renewal provided by a Continuous Variable Quantum Key Distribution (CVQKD) system. Our system was operational and able to encrypt point-to-point communications during more than six months, from the end of July 2010 until the beginning of February 2011. This field test was the first demonstration of the reliability of a CVQKD system over a long period of time in a server room environment. This strengthens the potential of CVQKD for information technology security infrastructure deployments.
\end{abstract}


\section{Introduction}
\label{sec:intro}

Quantum Key Distribution (QKD) \cite{sca:rmp09} is among the first industrial applications of the field of quantum information processing. Its natural commercial target is network security, since this technology allows two distant parties to share a secret key through the exchange of quantum states even in the presence of an eavesdropper, provided that the parties share an auxiliary authenticated classical communication channel. Contrary to all known classical schemes, the security of the established key can be proven without making any assumption on the capacities of the eavesdropper (for example computational power, knowledge of efficient algorithms, amount of memory). In theory, this key can be combined with an information-theoretically secure encryption method, the one-time pad (OTP), which requires a key that has to be as long as the message. However, the latest long-term field demonstrations, the Tokyo QKD network \cite{sas:oe11} and the SwissQuantum network \cite{stu:njp11}, report a secret key rate lower than 1 Mbit/s, which makes OTP incompatible with most of practical applications that require key rates above 1 Gbit/s. If high bit rates are required, a practical solution is to use the keys issued from QKD to renew keys used in classical symmetric algorithms like the FIPS Advanced Encryption Standard (AES) \cite{AES}. Since each QKD key is completely independent of keys generated earlier, renewing keys forces an attacker to perform a new attack to obtain the key after each renewal. This \emph{forward secrecy} property cannot be achieved with classical symmetric schemes. It can, however, be achieved with classical asymmetric schemes but only under some complexity assumptions \cite{kun:arxiv11, ioa:arxiv11}.

In order to become an essential part of current network infrastructures, QKD systems have to pass integrability and reliability tests. Systems that rely on encoding the information on discrete variables, such as the phase or the polarization of single photons, have been widely tested. Commercial products based on such systems have been developed by ID Quantique \cite{IDQ} and MagiQ Technologies \cite{MAG}. AES key renewal was demonstrated in \cite{era:njp10}, where the ID Quantique QKD system was combined with an AES-based encryptor allowing to encrypt 1 Gbit/s communications, while the long-term reliability of this technology was tackled in \cite{stu:njp11}. In comparison with QKD based on discrete variables, continuous-variable QKD (CVQKD), relies on encoding the information on continuous variables such as the quadratures of coherent states \cite{gro:prl02}, that has been implemented in a great variety of situations \cite{gro:nat03, lan:prl05, lor:apb04, lor:pra06, lod:pra07a, tok:cle07, qi:pra07, sym:pra07, fos:njp09, els:njp09, din:opex09, hei:apb10, sym:opt10, shen:pra10} (for a review of quantum information with continuous variables see \cite{wee:arxiv11}). This has important practical advantages: the homodyne detection hardware does not require any specific component, such as actively cooled single-photon detectors, and exhibits a better compatibility with a Wavelength Division Multiplexing (WDM) environment \cite{qi:njp10}. As recently created companies, Quintessence Labs \cite{QUI} and SeQureNet \cite{SQN}, pursue the development of a new generation of CVQKD technologies, it is imperative to demonstrate the integrability and reliability of CVQKD in a long-term field deployment.

We report here on the design and performance of the Symmetric Encryption with QUantum key REnewal (SEQURE) project demonstration. In the context of this project, a point-to-point classical symmetric encryption link using keys provided by a CVQKD system was installed in a production environment and ran during six months. This is the first demonstration of the long-term stability of CVQKD.

\section{SEQURE demonstration field test}
\label{sec:fieldtest}

\subsection{Structure of the demonstration}
\label{sec:demo}

\begin{figure}
\centering
 \includegraphics[width=120mm]{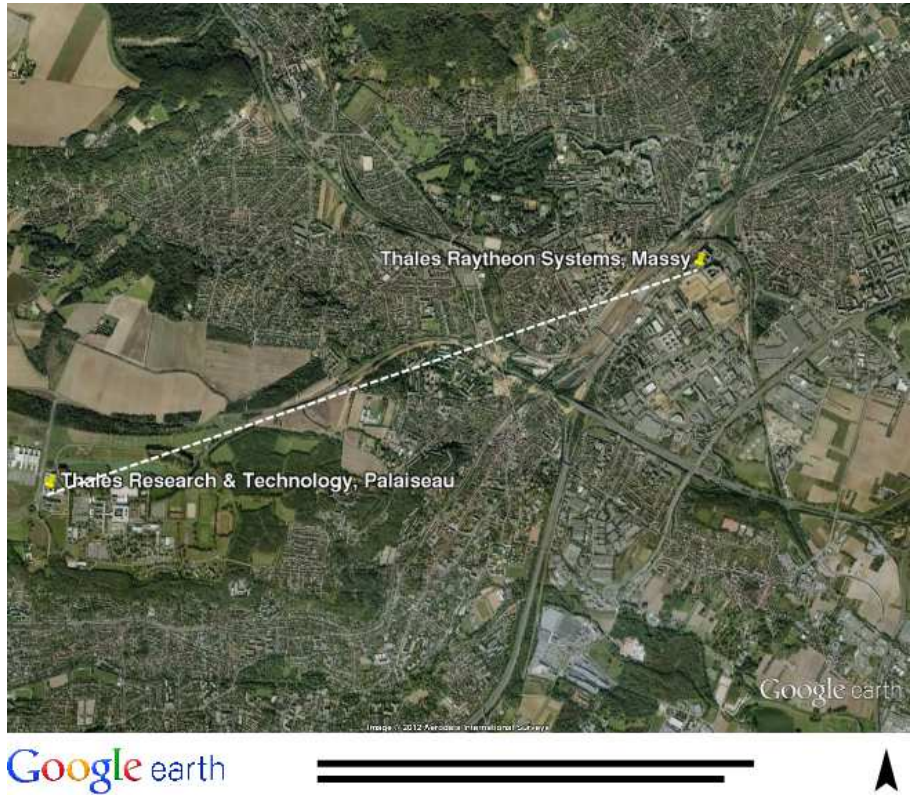}
  \caption{Map of the SEQURE demonstration. The two nodes are located in the cities of Massy and Palaiseau in the southwest of Paris.
The dashed line shows the 5 km straight path between the two sites, whereas the actual length of the fiber is 17.7 km. \copyright Google Maps - 2012}
 \label{figure:map}
\end{figure}
\begin{table}[t]
 \centering
 \begin{tabular}{c c}
  \hline
  Length of fibre (km) & Optical loss (dB)\\
  17.7 & 5.6\\
  \hline
 \end{tabular}
 \caption{SEQURE demonstration link characteristics.}
 \label{tab:link}
\end{table}

\begin{figure}
\centering
 \includegraphics[width=120mm]{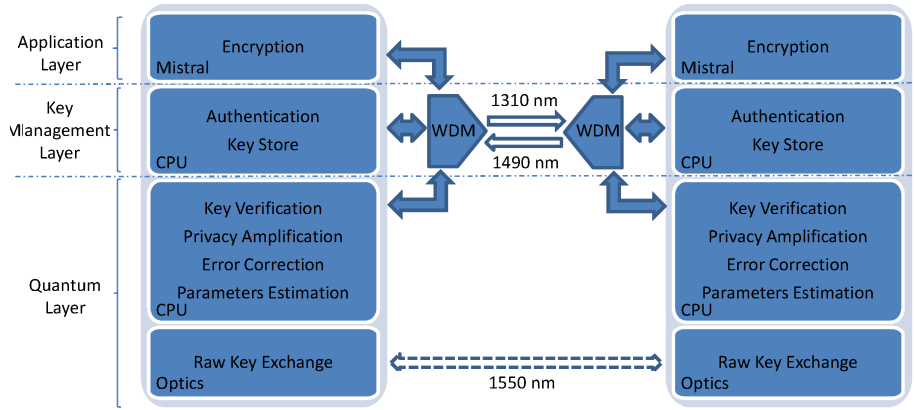}
  \caption{Layer structure of the SEQURE demonstration. The dashed bidirectional arrow represents the fibre used for the Local Oscillator (LO) and the quantum signal,
while the plain arrow stands for the wavelength multiplexed classical signals: the reconciliation part of the CVQKD, the key management layer and the encrypted traffic.}
 \label{figure:layers}
\end{figure}

We first discuss several important features of the SEQURE demonstration. The demonstration involves two nodes located in:
\begin{itemize}
 \item Palaiseau (Thales Research \& Technology France)
 \item Massy (Thales Raytheon Systems)
\end{itemize}
The characteristics of the link are summarized in table~\ref{tab:link} and a map is given in figure~\ref{figure:map}. For the generation and management of the secret keys, we used the layered architecture (see figure~\ref{figure:layers}) that was employed in the SECOQC FP6 European project network \cite{peev:njp09}. Note that the Tokyo QKD network \cite{sas:oe11} and the SwissQuantum network \cite{stu:njp11} also used the same architecture. Its main feature is that a layer of abstraction, the key management layer, is added between the physical medium used for the QKD and the application layer that uses the produced keys to secure classical communications. Then, it is straightforward to replace a QKD technology with another and, in our case, to extend the SEQURE demonstration point-to-point setup to a multipoint configuration.

In more detail, the SEQURE demonstration layers are the following:
\begin{itemize}
 \item a quantum layer implemented with a CVQKD point-to-point link (developed jointly by Thales Research \& Technology and Institut d'Optique/CNRS \cite{lod:pra07a, fos:njp09})
 \item a key management layer allowing to authenticate the reconciliation traffic of the quantum layer and to provide symmetric keys to the application layer (this software was developed by the Austrian Institute of Technology (AIT, formerly Austrian Research Center ARC) during the SECOQC project \cite{peev:njp09})
 \item an application layer where the keys coming from the key management layer are used by end users to encrypt their communications with a classical symmetric encryptor Mistral Gigabit (developed by Thales Communications)
\end{itemize}

\begin{figure}
\centering
 \includegraphics[width=120mm]{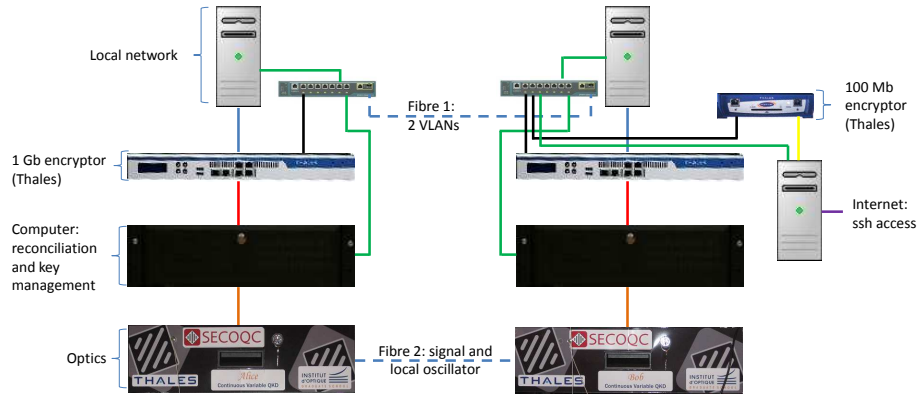}
  \caption{Structure of the SEQURE demonstration. The dashed lines correspond to the two fibres. Fibre 1 is used for multiplexed classical communications, fibre 2 transmits the physical pulses that are used to establish the raw key. The colors correspond to the different types of traffic: blue is plain text, black is the encrypted traffic VLAN, red is key renewal, orange is optics control and raw key traffic, yellow is the control of Mistral products configuration, purple is the Internet link, green is the monitoring traffic VLAN.}
 \label{figure:physical}
\end{figure}

The physical layer of the link is composed of one pair of dark fibres. One fibre is used as a quantum channel \footnote{Note that in a CVQKD system there are two time-multiplexed physical channels on the optical fibre since a classical signal, namely the local oscillator, is transmitted on the same fiber as the quantum signal.} and the other one is used to transmit all the classical channels. In the SEQURE demonstration there were several types of classical channels:
\begin{itemize}
 \item the channel for the reconciliation protocol of the QKD system
 \item the channel for cryptographic applications
 \item the channel for the monitoring of all the devices
\end{itemize}
A diagram of the components of the system is shown in figure~\ref{figure:physical}.

Since all the above classical channels need to operate in both directions, they are multiplexed using WDM techniques. The wavelengths used are 1490nm (uplink) and 1310nm (downlink). Standard GigaBit Interface Converters (GBIC) were used to convert ethernet optical signals propagating in fibres into ethernet electric signals propagating in copper cables. The optical fibres carrying the classical channels exhibited significant losses in the L-band; therefore Small Form-Pluggable (SFP or Mini-GBIC) modules specified for a 40-km link (Prolabs GLC-BX-D/U) were used. These modules were inserted into 8-port Cisco switches (WS-C2960G-8TC-L) with one dual port. The different classical channels were realized using only one classical link and multiplexing via Virtual Local Area Networks (VLANs). One VLAN is used for the reconciliation and the encrypted traffic, another VLAN is used to monitor all the devices.

The VLAN developed to monitor the demonstration was connected to a Management Center located in Thales Research \& Technology in Palaiseau. Remote accessibility to this management center was provided by secure shell (ssh) connections allowed only for legitimate users.

In case of power cuts, rack-mounted remote-control power switches (ePowerSwitch) could be used. They provide a secure web server interface allowing to switch on and off specific devices.

\subsection{The quantum layer}
\label{sec:quantum}

\begin{figure}
\centering
 \includegraphics[width=120mm]{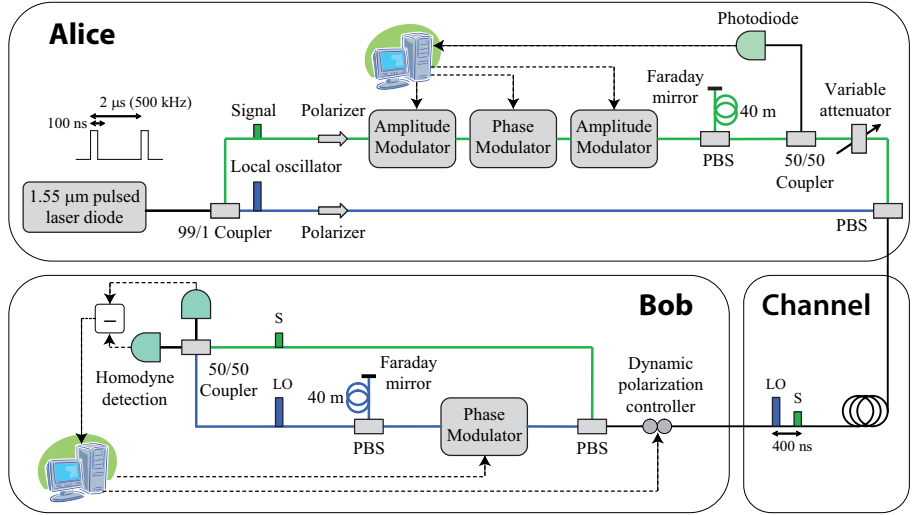}
  \caption{Optical layout of the CVQKD prototype. Alice sends to Bob 100 ns coherent light pulses generated by a 1550 nm telecom laser diode pulsed with a frequency of 500 kHz. These pulses are split into a weak signal and a strong local oscillator (LO) with an unbalanced coupler. The signal pulse is modulated with a centered Gaussian distribution using an amplitude and a phase modulator. The variance is controlled using a coarse variable attenuator and a finely tuned amplitude modulator. The signal pulse is 400 ns delayed with respect to the LO pulse using a 40 m delay line and a Faraday mirror. Both pulses are multiplexed with orthogonal polarization using a polarizing beamsplitter (PBS). The time and polarization multiplexed pulses are then sent through the channel. They are demultiplexed on Bob's side with another PBS combined with active polarization control. A second delay line on Bob's side allows for time superposition of signal and LO pulses. After demultiplexing, the signal and LO interfere on a shot-noise limited balanced pulsed homodyne detector. A phase modulator on the LO path allows for random choice of the measured signal quadrature.}
 \label{figure:hardware}
\end{figure}

The quantum link is composed of a pair of optical devices, whose hardware description is given in figure~\ref{figure:hardware}. This is a one-way implementation, where Alice sends to Bob 100~ns coherent light pulses generated by a 1550 nm telecom laser diode pulsed at a frequency of 500 kHz. These pulses are split into a weak signal and a strong local oscillator (LO) with an unbalanced coupler. The implemented protocol uses Gaussian modulation of coherent states \cite{gro:prl02}: the signal is randomly modulated following a centred Gaussian modulation in both quadratures, using an amplitude and a phase modulator. The random numbers used for this modulation are provided by Quantis, a physical Random Number Generator (RNG) from ID Quantique \footnote{http://www.idquantique.com/true-random-number-generator/products-overview.html}. The signal pulses are then attenuated roughly by a variable attenuator and finely by a second amplitude modulator, allowing to control the variance of the Gaussian distribution exiting Alice's device.

The signal and LO are then transmitted through the optical fiber without overlap using time and polarization multiplexing. One 400~ns delay line, composed of a 40-m single-mode fibre followed by a Faraday mirror, is inserted into Alice's signal path for the time multiplexing. Polarization multiplexing is achieved by a polarization beam splitter (PBS) on Alice side. Both pulses propagate through the fibre with orthogonal polarizations and a 400~ns time delay. They are demultiplexed on Bob's side with another PBS combined with active polarization control. A second delay line on Bob's side allows for time superposition of signal and LO pulses.

After demultiplexing, the signal and LO interfere on a shot-noise limited balanced pulsed homodyne detector (HD). The electric signal coming from the HD is proportional to the signal quadrature $X_\phi$, where $\phi$ is the relative phase between the signal and the LO. Following the protocol, by applying a $\pi/2$ phase shift, the phase modulator on Bob's LO path allows one to measure randomly either $X_0$ or $X_{\pi/2}$.

Finally, feedback controls are implemented to allow for a stable operation of the system over several months. Polarization drifts occurring in the quantum channel are corrected using a dynamic polarization controller that finds an optimal polarization state at the output of the channel. Temperature drifts affect lithium niobate, the active material used in the amplitude and phase modulators, therefore the voltages that need to be applied to reach the target modulation vary with temperature. The photodiode on Alice's signal path is used for the feedback control of the amplitude modulators while the HD output is sensitive to phase and can be used to control the phase modulators.

It is important to note that the Gaussian modulation used in the implemented protocol \cite{gro:prl02} maximizes the mutual information between Alice and Bob, thus offering an optimal theoretical key rate against either individual \cite{gro:nat03} or collective \cite{gar:prl06, nav:prl06} attacks. However, it is hard to reconcile correlated Gaussian variables with low signal-to-noise ratios (SNRs). The limited efficiency of the error-correcting codes (typically 0.90 bit extracted per bit theoretically available) results in a limit of the secure distance in the order of 30 km in our case. However, new ideas have been proposed \cite{lev:pra08} and recently implemented \cite{jou:pra11} to increase the secret bit rate and sequre distance, still keeping the Gaussian modulation which has presently the most robust security proofs \cite{gar:prl06, nav:prl06}.

With respect to the classical communication, four steps are required (see figure~\ref{figure:layers}). First, a Parameters Estimation (PE) step is needed to compute estimates of the physical parameters linked to the exchange of quantum states through the quantum channel. These parameters are the modulation variance $V_A$, the transmission of the quantum channel $T$, and the excess noise $\xi$. For some measured transmission and excess noise the modulation variance $V_A$ is adjusted in order to optimize the secret key rate for a set of pairs (SNR, $\beta$) (where SNR is the Signal to Noise Ratio and $\beta$ is the efficiency of the error-correction procedure) corresponding to the set of available error-correcting codes \cite{lod:pra07a}. The other parameters used to compute an estimate of the secret information that can be extracted from the shared data, the electronic noise $v_{el}$ and the efficiency of the homodyne detection $\eta$, are measured during a calibration procedure that takes place before the deployment of the system and that is assumed to be performed in a secure environment. For the SEQURE demonstration, the second step, which is the error correction procedure, was based on a multilevel reconciliation algorithm using Low Density Parity Check Codes (LDPC). This data reconciliation algorithm is explained in detail in \cite{lod:pra07a}. The amount of data revealed during this step is subtracted from the secret information previously computed. The privacy amplification step described in \cite{lod:pra07a} allows us to extract the secret information from the identical strings shared by Alice and Bob after the error correction procedure. Finally, a key verification step ensures with an overwhelming probability that Alice and Bob secret keys are identical. This is simply done by revealing a small part of the final bits chosen at random.

\section{Security considerations}
\label{sec:security}

The authentication of the classical channel needed for the QKD protocol is performed by the cryptographic engine provided by the AIT software. A point-to-point authenticated channel is created by the Q3P protocol. It is based on the Wegman-Carter scheme \cite{weg:jcss79, weg:jcss81}. This authentication protocol, like other QKD implementations, requires an initial common secret. 

As for other families of QKD systems, some attacks can be implemented on a CVQKD system exploiting the imperfections of the setup. For example, the presence of excess noise, which is noise un excess of the shot noise, opens the possibility for partial intercept-resend attacks as demonstrated in \cite{lod:prl07}. This is why the shot noise level on the receiver side must be precisely known. Monitoring the physical parameters of the channel allows to upper-bound the information available to Eve. An efficient way to perform quantum hacking (see \cite{sca:rmp09}) on a QKD system consists in exploiting side-channels. In our setup, a linear relationship between the LO level and the shot noise is determined during the system calibration. Then the LO level is continuously monitored with one photodiode of the HD and the shot noise level is computed with the help of the previously calibrated relationship. It is used to convert in shot noise units all the physical quantities needed to compute the amount of generated secret data. Generally, the LO, which is a classical signal that can be manipulated by an eavesdropper, is a potential vulnerability \cite{fer:iqec07}. Monitoring the LO level is a counter-measure to such attacks.

\section{Performance of the quantum layer}
\label{sec:performance}

\subsection{Events}
\label{sec:events}

The system was stable and ran continuously during more than 6 months, from the end of July 2010 to the beginning of February 2011. The optical part did not require any human intervention during the full period of the demonstration. We list below the most significant problems experienced during the demonstration:
\begin{itemize}
 \item September 23 to September 29: the motherboard of Alice's computer in Massy failed and had to be changed (these two dates correspond to the two first marks on figure~\ref{figure:excess_noise_key_rate} and figure ~\ref{figure:key_128}). Error correction is the most demanding task in terms of computing power and is performed on Alice's side. 
 \item October 1 to October 31: the server room in Massy (Alice's side) was unavailable so the experiment had to be interrupted until it was started again in a new location (these two dates correspond to the two last marks on figure~\ref{figure:excess_noise_key_rate} and figure ~\ref{figure:key_128}).
 \item November 1: the system was restarted but the experimental conditions became continuously changing because of a lack of thermal regulation. However, the results could still be exploited.
\end{itemize}

\subsection{Excess noise}
\label{sec:excess}

\begin{figure}
\centering
 \includegraphics[width=120mm]{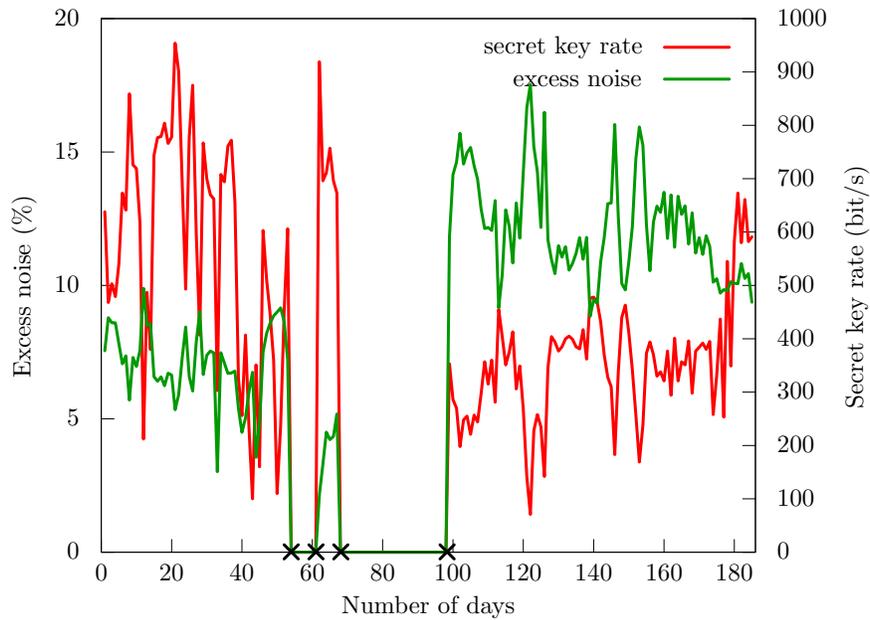}
  \caption{In red, the secret key rate during the SEQURE demonstration. In green, the measured excess noise during the SEQURE demonstration. During the first part (server room), this excess noise can be mainly attributed to the acoustic noise in the server room. In the second part, an additional excess noise occurred, that is attributed to thermal fluctuations due to the lack of thermal regulation in the room. The black marks correspond to the events listed in the section \ref{sec:events}.}
 \label{figure:excess_noise_key_rate}
\end{figure}

The excess noise was recorded during the full period of the experiment and is reported in figure~\ref{figure:excess_noise_key_rate}. On a daily scale, it is subject to variations linked to statistical fluctuations and experimental conditions like fibre vibrations. Keys are mainly produced with low values of the excess noise, while no keys are produced on blocks with a large excess noise because of the limited efficiency of the error correction scheme (about $90\%$, see \cite{lod:pra07a}). The system operation was rather stable during the 6 months but we can notice a significant difference in performance when the experiment at one site was transferred from the server room to the room with no thermal regulation. In fact, the excess noise obtained with the equipment in these degraded experimental conditions does not allow to obtain a positive secret key rate against collective attacks for the line transmission \cite{gar:prl06, nav:prl06}. As a result, a secret key rate against individual attacks \cite{gro:nat03} only was computed during the second part of the demonstration. This illustrates the importance of monitoring continuously the excess noise in order to evaluate the security of the keys \cite{lod:prl07}. It is important to note that this kind of problem is typical of an external environment and would not occur in laboratory conditions. Our system was still able to produce keys in those degraded conditions, although with an inferior performance. This illustrates the maturity of our setup.

\subsection{Secret key rate}
\label{sec:rate}

\begin{figure}
\centering
 \includegraphics[width=120mm]{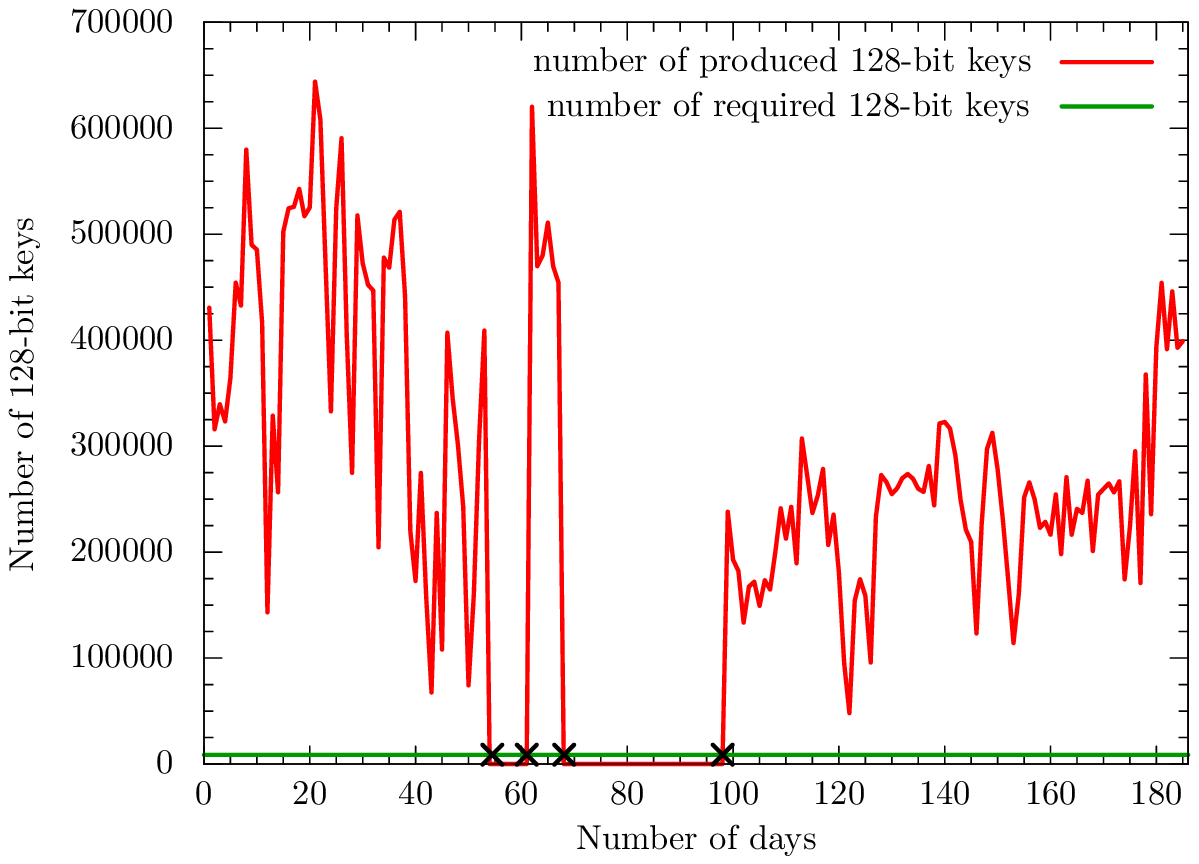}
  \caption{In red, number of 128-bit keys per day produced during the SEQURE demonstration. In green, number of 128-bit keys per day required for a key renewal every 10 seconds. The number of produced keys largely exceeds this limit. Before day 100, the keys were produced assuming collective attacks from an eavesdropper. After day 100, they were computed assuming only individual attacks because the excess noise was significantly higher. The black marks correspond to the events listed in the section \ref{sec:events}.}
 \label{figure:key_128}
\end{figure}

The keys generated by the quantum layer were used to refresh the Thales Mistral encryptors' 128-bit AES keys. The renewal period was 10 seconds, thus the quantum layer had to be able to generate 8640 128-bit keys per day, which is roughly 1 Mbit of key material. Then, a 20 bit/s secret key rate would be sufficient. This rate is much lower than the rate up to 2 kbit/s that our setup can produce with comparable line transmission and excess noise conditions \cite{lod:pra07a}. The ultimate performances of our system were obtained using a multithread data processing architecture with 2 cores devoted to the reconciliation and 1 core dedicated to the management of the hardware part \cite{fos:njp09}. In the present case only one core is required to perform the reconciliation, which results in an improved stability of the software and an improved stability of the overall system over long periods. Figure~\ref{figure:key_128} shows that the SEQURE demonstration was largely above this threshold. Figure~\ref{figure:excess_noise_key_rate} shows that the key rate was about 600 bit/s (calculated against collective attacks) during the first part of the demonstration and 400 bit/s key rate (calculated against individual attacks) during the second part.

\section{Performance of the encryption layer}
\label{sec:encryption}

Several tests were performed in order to ensure that the key renewal did not affect the operation of the encryptors. As no specific adaptation of the Thales Mistral products was performed for the project, it was clear that the encryptors could not deal with a key renewal period lower than 3 seconds. A 10 second period, as mentionned in the previous paragraph, was therefore chosen as a security margin. This period could seem arbitrary but it ensures that no more then $2^{35}$ bits are encrypted with the same key if we consider 1Gb/s data communications. This can be compared with the best known attack \cite{mat:eur93} on the former encryption standard, the Data Encryption Standard (DES), which requires $2^{43}$ plaintext - ciphertext pairs, that is $2^{49}$ bits of observed traffic with known plaintext.

Classical networking applications like big files transfers (more than 1 Gbyte), disk sharing and persistency of the network link were tested. In all cases, the performance of the Mistral Gigabit was not affected by the key renewal. 

\section{Conclusion and perspectives}
\label{sec:conclusion}

The SEQURE demonstration that we have presented shows that continuous-variable QKD can compare well with discrete-variable QKD with respect to robustness and reliability in a server room environment, whose operating conditions are harder to cope with than laboratory ones. Furthermore, it shows that CVQKD can be integrated easily with off-the-shelf network equipments such as symmetric encryptors as a part of a more complex network infrastructure. Integration into WDM networks could be also eased by tolerance of the CVQKD homodyne detection scheme to incoherent noise \cite{qi:njp10}. Moreover, if CVQKD WDM compatibility is confirmed in real optical network deployments, it will imply a significant decrease of the operational costs, which can stimulate further interest for this technology.

The operating distance of the implemented system can be improved by the recent developments of better error-correcting codes \cite{jou:pra11} without any hardware modification. These codes would also allow to produce keys secure against collective attacks even with the high values of the excess noise obtained during the second part of the demonstration. For distances higher than 100 km, the key management layer developed within the SECOQC project can still be used to share keys between two sites connected through several links.

As regards to the key rate, the current limitation of the system is not the optical part but the error correction speed which can be drastically improved using Graphics Processing Units (GPU) \cite{lod:pra07a, jou:prep12}. Furthermore, in order to take into account finite-size effects it is necessary to process large blocks ($\geq 10^8$ pulses) to extract the final key \cite{lev:pra10}. Then improving the error-correction speed allows to deal with finite-size effects without dramatically increasing the key production latency.

Finally, in a setting where QKD is used together with computational high-speed symmetric encryption like in SEQURE, it is not unreasonable to use a scheme based on minimal assumptions about the security of symmetric cryptography, like the Lamport signature scheme instead of using an initial secret key. This enables to initialize QKD with an exchange of \emph{authentic} values, which is easier to perform than an exchange of secret values \cite{kun:arxiv11}.

\section*{Acknowledgements}
We acknowledge support from the Agence Nationale de la Recherche under projects SEQURE (ANR-07-SESU-011), FREQUENCY (ANR-09-BLAN-410), Paris Region and ICT Cluster Systematic, and the ERANET project HIPERCOM. The authors thank Thales Raytheon Systems and Thales Research \& Technology for the loan of their server rooms. We thank France Telecom for making available two dedicated fibre links for the duration of the demonstration.

\end{document}